# Accessing site response through H/V ratio in Shillong City of northeast India

Rajib Biswas[1*], Saurabh Baruah[2]

1 Department of Physics, Tezpur University
   Tezpur 784028, Assam, India

2 Geoscience Division, CSIR North East Institute of Science and Technology
   Jorhat 785006, Assam, India

*Corresponding author: Email address: rajeeve_biswas2004@yahoo.com;
rajib@tezu.ernet.in

Abstract:

*We endeavor to assess site response exploiting ambient noise measurements which we carried out at 70 sites in different parts of Shillong City, one of the seismically active regions. We estimate the spectral ratio from these recordings. The spectral ratios reveal the resonant frequency for Shillong City in the range of 3 to 8 Hz. Short scale spatial variation in the resonant frequencies suggests that there is prominent lateral heterogeneity in the underlying layers. Besides, a non-linear earthquake site response analysis is also attempted using available geotechnical data. A good correlation is observed between site response analysis and HVSR results constrained by wide variation of resonant frequency at short distances. The results are found to be tallying well with the geological data.*

**Introduction**

Over the recent years, earthquake damage has become a real big concern for all including seismologists, civil engineers and policy makers. Although it is a well-known fact that earthquakes never cause causalities but structures do. These structures are basically located on different kinds of soils. Consequently, shaking due an earthquake experienced by them varies sometimes at short distances (Komatitsch and Vilotte, 1998; Ouchi et al., 2001), which is directly linked to damage (Cara et al., 2008; Bonnefoy-Claudet et al., 2008). These damages are attributed not only to magnitude of earthquake and its epicentral distance, but also due to local site effects which are essentially frequency dependent. These site effects arise chiefly because of topography and geology of site (Badrane et al., 2006). The geological structure behaves as a selective filter shaping spectrum of energy shaking the ground (Fernandez et al., 2000). The site effect is typified by resonance frequency and the associated ground motion amplification. Utilization of microseisms or ambient noise records have been proven to be an effective tool for determination of fundamental resonant frequency (Mundepi et al., 2009; Lombardo and Rigano, 2007). It is well known that soil deposits amplify ground motion. The amount of amplification depends on several factors including layer thickness, degree of compaction and age (Siddiqui, 2002). Amplification of seismic waves by sediments is an important parameter in determining seismic hazard. Noteworthy is the fact that the younger sediments amplify more compared to competent soil or bedrock. It is also a matter of concern that most of the urban settlements have grown along river valleys over such young and surface deposits (Hasancebi and Ulusay, 2006). This is because at sites, having soft soil or topographic and basement undulations, seismic energy gets trapped, leading to amplification of shaking of manmade structures.

One of the many reasons for choosing ambient noise is that it allows the quick & reliable estimate of site characteristics of any type of area, especially in urban environments in particular (Cara et al., 2008; Mundepi et al., 2009; Lombardo and Rigano, 2007; Duval et al., 2001; Lebrun et al., 2001; Panou et al., 2005; Gueguen et al., 2000; Garcia-Jerez et al., 2007; Cara et al., 2003). Apart from being a cost effective measure, it reduces time compared to estimating site characteristics from earthquake which has always been a time consuming as well as expensive process so far as the maintenance of equipment & man power is concerned.

When the frequency of oscillation of a site matches natural frequency of man-made structures, there will be drastic rise in amplitude which leads to damage of the later owing to resonance effects. So far estimation of PGA is concerned, many researchers show that PGA in near-field zone is not amplified on soft ground and is even slightly decreased. For example, during the San Fernando earthquake for epicentral distances >50 km, PGA is larger on soil while for <50 km, it is smaller (Duke et al., 2007). Simultaneously, it is observed that the most important factor for relation between horizontal and vertical components is the vibration level for strong ground motion records (Mikhailova and Aptikaev, 1996). Therefore, investigation of each site condition is an important step towards earthquake hazard estimation. One of the most striking features of Northeastern India is that most of its cities and densely populated settlements are located in valley, sedimentary basins or hills etc. The complex geotectonic setup of Shillong City needs better understanding in the light of site-amplification characteristics. The city in particular has not been covered by any proper estimation of fundamental frequency and related parameters.

In this study, we try to shed light upon the site characteristics of Shillong City in terms of resonant frequency, site amplification etc. with adoption of spectral ratio, i.e., H/V ratio methodology (Nakamura, 1989; as modified by Bard, 1999). Further, an attempt has been made to simulate ground motion in this area w.r.t felt earthquakes with the help of nonlinear earthquake site response analysis.

**Geological and Seismotectonics of Shillong City**

Shillong, the capital of Meghalaya, NER, India, is situated in an almost elliptically shaped Shillong Plateau (SP). With an area coverage of 6430 square kilo meter, the population of Shillong City is rising (≈2, 70,000) which catapults rapid urbanization. As a result, man-made structures especially high rise buildings are being built & under construction as well in the central part of the township.

Shillong has an altitude of 1496 meters above sea level. It is bounded by the Umiam Gorge to the North, Diengiel Hills in the Northwest and Assam valley to the north-east. The SP with an Achaean gneissic basement and late Cretaceous-tertiary sediments along its southern margin is bounded by the Brahmaputra graben in the north and by Dauki fault in the south (Rao et al., 2008) Existence of several lineaments in and around city having major trends along NE-SW, N-S and E-W directions (Chattopadhaya et al., 1984) makes the city more vulnerable so far earthquake hazard is concerned.

Shillong forms the type area of Shillong series of parametamorphites, which includes mostly quartzites and sandstones followed by schist, phyllites, slates etc. (Sar, 1973; Kalita, 1998). The base of Shillong series is marked by a conglomerate bed containing cobbles and boulders of earlier rocks, i.e., Archaean crystalline, which formed the basement over which the Shillong series of rocks were originally laid down as sedimentary deposits in Precambrian times, probably in shallow marine conditions. The rocks that were intruded by epidiorite rocks are known as "Khasi Greenstone" as depicted in the Geological map of Shillong in Fig. 1. The Khasi Greenstone (Rao et al., 2009; Bidyananda and Deomurari, 2007) is a group of basic intrusives in the form of linear to curvilinear occurring as concordant and discordant bodies within the Shillong group of Rocks suffered metamorphism (Srinivassan et al., 1996). These kinds of rocks are widely weathered and the degree of weathering is found to be more in topographic depressions than in other areas. The metabasic rocks are more prone to weathering than the quartzite rocks. In low lying areas, valley fill sediments are also prominent.

The study area, Shillong City is a component of Shillong Plateau which is regarded as one of the most seismically active region in NER, India. Northeast Region (NER) of India, bounded by latitude ($28-30^0N$) and Longitude ($89-98^0E$), is seismically one of the most active zones in the world where sixteen large (M>7.0) and two great earthquakes of June 12, 1897(M 8.5) (Oldham et al. 1899) and August 15, 1950 (M=8.7) (Poddar, 1950) occurred during the last hundred years. These two great earthquakes have caused extensive damage killing a total of 3,042 lives and a total loss of $ 30 million (Tillotson, 1950). In this context, the NER is categorized as seismic zone V of India (BMTPC, 1983). Shillong Plateau belonging to NER, India is a part of the Indian Shield, which is separated out from the peninsular shield and moved to the east by about 300 Km along the Dauki Fault (Evans, 1964). The gigantic E-W trending Dauki Fault separates the plateau from the Himalaya to the North. The E-W segment of the river at the northern boundary of the Plateau is named Brahmaputra fault (Nandy, 2001). The Plateau is separated out from the peninsular shield and moved to the east by about 300 km along the Dauki fault (Evans, 1964). It is surrounded by many small & large faults & lineaments. Towards the western part of Shillong, there exists the active Barapani Shear Zone. It is one of the major thrust faults prevailing in this region. The complex geodynamics of Shillong Plateau resulted in the Great Assam earthquake of 1897. As reported by Bilham et al., (2001), this ($M_S=8.7$) earthquake caused severe damage to the settlements of this area (Shillong city), causing causalities in a large dimension. It may be

mentioned about a significant earthquake of June 1, 1969 with a magnitude of 5.0 having an Epicentral distance of 20km from Shillong (Gupta et al., 1980) which was felt rigorously. According to Gupta, since 1970, there has been gradual decrease in P-wave velocity yielding a speculation that the region is experiencing a dilatancy stress precursory to a large earthquake. According to Khatri et al., (1992), the Shillong Massif shows a pertinent seismic activity with an average of 10-15 small magnitude earthquakes per day. Over the past hundred years, there were instrumental records of 20 large earthquakes. With the advent of seismic networks set up by NEIST Network, IMD, there has been a tremendous improvement in the recording of micro tremors. The moderate magnitude seismicity in the Shillong area is somewhat confined to Barapani Shear zone. During the past recent months, there has been a contemporary rise in the no. of felt tremors whose epicenters lie within the vicinity of Shillong City. In the context of this ongoing pattern of seismicity, the current study will be an initiative towards microzoning of this region. The major tectonic features surrounding Shillong is displayed in Fig. 2.

**Field survey and data acquisition**

The ambient noise survey was carried out in Shillong City covering 70 sites in total. Being densely inhabited city, there are certain constraints regarding noise recording in urban environments. This resulted in less no. of data points. Besides, being a hilly terrain, the ambient records could not be accrued uniformly. To ensure reliable noise recording, the survey was performed multiple times where there was a possibility of artificial noise sources such as moving cars, pedestrians. Moreover, quiet environment & good weather condition had been our prime requisite during data acquisition process.

During data acquisition, the sensors were installed at the recording sites strictly following the Guidelines by SESAME, 2004. A three component short period S-13 seismometer from M/S

Teledyne Geotech, USA having natural period of 1s was used. The duration of data recording was more than 1 hour. We utilized 24 bit Reftek 72-0A digitizer. The recordings were digitized at 100 samples per second and recording time was maintained by Reftek GPS clock. The GPS locations of the ambient noise sites are shown Fig. 3.

**Data Processing**

Data from each site have been processed using LGIT Software, i.e.; SESARRAY package. Since we adopted identical instrumentation for all three components, no instrumental correction has been applied. With proper stable noise free from transients in window span of 30-40s range, a 5 % cosine taper was applied on both sides of the window of the signal of vertical (V), North-South and East-West components. It was then followed by FFT of the three components to obtain the three spectral amplitudes with a smoothing factor of~40 (Konno and Ohmachi, 1998).

As per SESAME Guidelines (2004), the reliability of H/V peaks were checked with appropriate standard deviations. We did another crosscheck of these peaks by subjecting them through Randomdec method (Huang et al., 1999; Dunand et al., 2002; Haghshenas et al., 2005) to ensure whether an H/V peak is of natural or anthropic origin. As observed, the origin of the H/V peak is ascertained to be of anthropic origin if the critical damping is found below 5%; otherwise it is considered to be generated by natural origin wherein we get a critical damping above 5%. This is exemplified in Fig. 4 [site 42].

**H/V Results**

The H/V ratios are evaluated for the sites where ambient noise recordings were made. Particularly, two kinds of frequencies are observed to exist from the H/V ratio estimates. The first category of frequency is ranged between 1 to 2 Hz while the other range of frequency is found above 2 Hz, as displayed in Fig. 5. The low frequency peaks as illustrated in Fig. 6 might be related to the existence of stratum with soft sediments. Similarly, the resonant frequencies which are placed in higher category of frequency might be caused by the presence of hard soil strata. These results are more detailed in our next observation where we aim at searching for correlation with local geology.

Reasonably, sharp peaks are observed in the Site 30, 42, 13 & 64. These sites are representative of the H/V ratio estimates performed for the majority of the sites. The sites portrayed in Fig. 6 yield some low frequency peaks which are not sharply defined.

Several theoretical 1D investigations (Field et al., 1993; Tokeshi et al., 1998; Lermo et al., 1994; Wakamatsu et al., 1996) show that H/V ratio exhibits fundamental frequencies with sharply defined peaks when there exists a sharp impedance contrast between the surface layer & the underlying stiffer formations. Comparison of Fig. 5 & Fig. 6 reveal the fact that the low frequency peaks, which are not so sharp, might be caused by layers having moderate impedance contrast with the underlying formations. The overall pattern of fundamental frequency from H/V ratio estimates corresponding to 70 sites of Shillong is depicted in the form of contour in Fig. 7. It is observed in the contour that certain pockets which are encompassed by higher frequencies from 5 to 7 Hz are covered by compact strata of rocks such as greenstones, quartzite as evidenced in the Geological map of Shillong (Fig. 1). Simultaneously, certain sites in the heart of the Shillong city exhibit lower fundamental frequency in the range of 1 to 2 Hz. When correlated with geological map, it yields that these sites, exhibiting lower fundamental frequency, are marked by presence of weathered soil cover. This well substantiate the findings of low fundamental frequency.

**H/V profiling-relation between amplitude of the frequency w.r.t sites**

H/V profiling is an important feature to visualize how the H/V ratio estimates vary over a certain section. This allows one to have a general inspection regarding the variation of resonance frequency in accordance with distance of the sites from each other. Additionally, the corresponding amplifications levels at the site can also be studied with this. In this study, two profiles are considered in order to observe the variation with respect to resonance frequency or distance. This has been done keeping in view of the close proximity of the borehole logs with that of the ambient noise recording sites. AB and CD are the two profiles wherein the variation of H/V amplitudes is observed, as shown in Fig. 3. Fig. 8 illustrates the H/V profiling corresponding to the profile AB. The profile is characterized by comparable variation in frequencies, covering very low to very high frequencies. The sites numbered 4, 6 and 67 exhibit higher frequencies in the range of 5.5 to 8 Hz. Lower frequencies less than 2 Hz are observed in two sites numbered 2 and 50. Another aspect can also be contemplated in this figure. The sites showing higher range of frequencies are also masked by higher amplitudes. As for the CD profile, the variations are observed in Fig. 9. It is quite evident in the figure that except two sites, majority of the sites correspond to higher frequencies. It is found to be in the range of 4 to 6 Hz. The two sites numbered 22 and 51 refer to low frequency sites. So far the amplitude variation corresponding to the peak frequencies are concerned, similar trend is visible as observed in AB profile. Though the level of amplitudes is quite high compared to the former, relatively higher frequencies encompasses the higher amplitudes. All these observations are substantiated in the next section when they are correlated with the existing litholog available in the immediate neighborhood of the recording sites.

**Correlation with borehole data**

There was a dearth of litholog information of Shillong City. However, we could accrue litholog information of 10 sites exclusively in Shillong City. In order to observe any correlation between fundamental frequency and geology, we focused on two specific profiles AB and CD in synchrony with H/V profiling as well as in conformity with the location of the boreholes & the neighboring ambient noise recording sites. Profile AB aligns with the three borehole sites available in the region. Similarly, profile CD which trends in NE direction covers six borehole sites. With the objective of correlating the existence of fundamental frequencies with existing lithology, many researchers resorted to reconstructing the geology with the available geophysical information. Mention may be made about the works of Cara et al. (2008); Hasancebi et al. (2006). On the basis of borehole data, we made an attempt to reconstruct geological cross-sections along the two observed profiles as displayed in Figs. 10 & 11, following Cara et al. (2008).

The resonance frequencies estimated from H/V ratio in different sites are correlated with reconstructed geological cross-section. As shown in Fig. 10, we divide the resonance frequencies into three different ranges. The first range is from 1 to 3.5Hz; next 3.5 to 5.5Hz & the highest one range between 5.5 to 7.5Hz or above. Similarly pertaining to profile AB oriented along NW-SE, there is a transition of resonance frequency from higher in northwest part to lower frequency towards southeast region. In the higher frequency portion, it comes within an average of 5 Hz. Thus, it can be attributed that the southeast part is dominated by lower resonance frequency in comparison to northwest part.

The existence of higher frequencies may be attributed to the presence of hard rock strata along the NW part of the profile. This observation holds well as inferred from the site specific ambient noise analysis except some non-conformity in site no. 28 & 29. Notably, site no. 4 & 5 showed higher resonant frequencies, consistent with the presence of more compact thicker rock strata, as evident in Figure 1. Towards the SE edge of the profile, mainly low frequency dominates the region. An inspection at the reconstructed portion leads to the fact that there exists a comparatively lower density clay stratum. The little variation may be attributed to the degree of compactness of the strata as also inferred in Cara et al., 2008.

However, a wide variability in resonance frequencies starting from SW part to N-S region is contemplated in profile CD, as illustrated in Fig. 11. Along the profile, site no's 69, 56 & 39 show a gradual transition to higher fundamental frequencies from the lower resonant frequency characterizing the sites 60, 61 & 62. This might be attributed to the presence of low S-wave velocity zones beneath. Perhaps, thickening of weathering layer can be another contributing factor towards this abrupt transition. Additionally, it is observed that sites 37, 32, 34 & 21 exhibits lower fundamental frequencies. Towards SE direction along the profile there is a sudden transition to higher value of fundamental frequency observed as 6.5 Hz at site no 20. This sudden transition may be attributed to the local geology of the site which is the juncture of sand granite and sand rock as envisaged from the geological map (see Fig. 1). In both these constructions, one particular aspect was the undulating topography existing in the two profiles.

**Non-linear site response analysis**

In order to validate the H/V estimates, we perform a numerical analysis to simulate one felt earthquake recorded by local Strong motion Networks. The hypocentral parameters are given in Table 1. The reason for choosing this earthquake lies in the fact that it has the epicenter situated at a distance of less than 140 km. This earthquake of magnitude 5.1(Mw) was felt at several places in the adjoining areas of Shillong City. With a view to simulate this earthquake, we adopted the technique of nonlinear earthquake site Response analysis (NERA) developed by J.P Bardet and T. kBobita, (2001). Utilizing the available N values as input parameter, ground acceleration of the 19th August, 2009 event could be simulated at ten Borehole sites.

It has become a wide practice of using empirical relation that relates shear wave velocity with N values of Standard Penetration Test when there is a dearth of Geophysical parameters. In this study, the empirical relationship of Imai and Tonochi, 1982 ; Ohta and Goto, 1978 are used to compute shear wave velocity in order to define the soil profile (see Fig. 12), required for the numerical analysis. The empirically determined estimates of shear wave velocities are found to be consistent when compared with a recent study by Raghukanth et al., (2009) in Imphal City, NER, India. Though the site geology is different for Imphal and Shillong City, it implies that this procedure of empirical estimation is valid in this study region also. For example, in Fig. 12, three soil profiles are displayed viz., Arbhutnot Road, Pynthorumlhrah and East-Pinewood Hotel. Here, the maximum shear wave velocities as found from the empirical relations come in the range of 450 m/s. While adopting the aforementioned empirical relations, we adhered to the depth range pertaining to the borehole sites. Subsquently, we intend to validate the resonance frequency estimated through non-linear earthquake site response analysis near the borehole sites with the resonance frequency obtained through H/V ratio from ambient noise. Utilizing the shear wave velocity profile, the soil profiles are estimated at all the 10 sites.

Simultaneously, the spectral acceleration and fundamental frequencies are estimated through nonlinear earthquake response analysis. Spectral acceleration bears a significant role in the damage associated with strong motions. It is referred to as the acceleration experienced by the existing structures whereas peak ground acceleration is the acceleration experienced by a particle. Fig. 12 exemplifies the estimated parameters for Umpling, one of the Borehole sites. The estimated values are shown in Table 2. The estimates indicate varying ground motion parameters pertinent to all the borehole sites.

Simulation of the accelerograms allows estimation of fundamental frequencies which range between 5-12Hz. The fundamental frequencies derived from nonlinear response analysis are found to be consistent with the estimations obtained from H/V technique in majority of the borehole sites as detailed in Table 3. In case of sites e.g. Basic Science, Sawlad & Near 13A tank, the resonance frequencies are overestimated w.r.t H/V ratio estimates. The conformity of the fundamental frequencies done from simulation of ground motion of the event with the H/V computations indicates the consistency in our approach towards estimating resonance frequencies.

Higher spectral acceleration is observed in Pynthorumkhrah which may be attributed to the weathered soil cover, as evidenced from local geology. Validation of our results through simulation of ground motion of earthquake with the help of Nonlinear earthquake site response analysis affirms the notion that H/V ratio can be utilized for a reliable estimation of fundamental resonance frequency.

**Discussion**

In this work, an attempt is made to estimate site response through horizontal to vertical ratio from ambient noise analysis. Pertaining to this aim, ambient noise records are utilized which were accrued in seventy different sites in Shillong City. The collection of ambient noise records at these sites were performed keeping in view the proper soil-sensor coupling and other data related parameters. Afterwards, a distinctive analysis towards selection of suitable data is done, followed by preprocessing based on the SESAME, (2004) guidelines. Site-specific resonant frequencies for all the seventy sites are computed through H/V ratio. Prior to this, the H/V curves are subjected to different reliability tests so as to decipher the reliable fundamental frequencies. Moreover, identification of the nature of origin of peak resonant frequencies is accomplished in order to achieve the best one pertinent to the site. All these are performed with the aim of inferring reliable site characterization of the study region in the context of resonant frequencies and site amplification behavior. Apart from this, influence of local lithology is also explored with a view to correlate with the computed frequencies. Besides, an empirical relationship between fundamental frequencies and sedimentary thickness pertinent to this region is developed. In one of our recent work, Biswas et al., 2015, the sub-surface shear wave velocity is empirically related to the depth of the overburden thickness. The findings, herein, well substantiate the results as found in Biswas et al., 2015.

So far the estimations regarding the H/V peak frequencies are concerned; two categories of site frequencies are contemplated. The higher side of fundamental frequencies observed in majority of the sites could be correlated with the compact form of strata. It could also be attributed to the thinning of sedimentary structures beneath the sites as observed in many of studies e.g., Lombardo et al., (2001); Guillier et al., (2004). The sites revealing lower frequencies can be interpreted to be characterized by overburden thickness. The overall estimate of fundamental frequencies in the form of contour (refer to Fig. 7) indicate a wide variation. Some sites (e.g Laban, Rinsa Colony) [Lat-$25.57^0$ to $25.59^0$; Long-$91.87^0$ to $91.89^0$] situated in the central part of Shillong City yield lower estimates of fundamental frequencies in the range of 1 to 2 Hz. This observation matches well with the local site condition. However, certain sites (e.g, Pynthorumkhrah, Lumparing etc.) in Shillong City reveal higher estimates of resonance frequencies. These observations are established to be in good agreement with the local geologies, as evidenced in the geological map which reveal the fact that the sites showing higher estimates of resonant frequencies are underlain by different form of rock strata such as sandstones, quartzites and conglomerates etc.

In order to better understand the spatial variation of fundamental frequencies in Shillong City, the subsurface geology along two profiles AB and CD pertaining to the availability of borehole information in conformity with the locations of the ambient noise sites are reconstructed (refer to Figs. 10 & 11). When correlated with the variation of the fundamental frequencies, the site geology is found to tally with the estimated peak frequencies through H/V ratio. Cara et al., (2009) carried out such type of correlation studies wherein the variability in H/V peak frequencies was interpreted in terms of the existing lithology. In both the two observed profiles AB and CD, the sections drawn by virtue of the borehole-log information indicate a heterogeneous formation of structures. In fact, it is well represented by the corresponding fundamental frequencies observed in these two profiles which also show a wide variability in response to the underlying formations characteristic of the reconstructed cross section. Along the AB profile, lower frequencies cover lower amplitudes whereas higher frequencies correspond to higher amplitudes. Souriau et al., (2007) observed deamplification at frequencies greater than 3 Hz. The variation of amplitude as found in the CD profile reveals a similar trend which can be interpreted to be caused by topographic effects (Dubos et al., 2003).

Through Numerical simulation of input acceleration histories aided by geotechnical inputs in the form of lithologs and N-values from Standard Penetration Test, it is sought to analyze site-specific non linear soil behavior. This necessitates the incorporation of shear wave velocities which are actually deduced by the widely used empirical relationships (Ohta and Gotto, 1978; Immai & Tonochi, 1982) exploiting the N-values. Here also, the estimates of the ground motion parameters particularly the peak ground acceleration shows an anomalous behavior pertaining to the borehole sites.

**Conclusion**

In this study, we endeavor to map the distribution of fundamental frequency obtained from recorded ambient noise with H/V ratio. In spite of some limitations of H/V method,–we obtain some notable features. Most importantly it is observed that fundamental resonance frequency for Shillong city ranges between 3 to 8 Hz. In fact, most parts of the area are characterized by higher fundamental frequency which may be correlated with the existing thicker strata of basement rocks underlying the surface layer yielding sharp peaks in the H/V ratio. At short distances, wide variation of resonance frequencies in the H/V ratio is observed. This may be viewed as an indication of lateral heterogeneity prevailing in soil layers of the region. Moreover, a good correlation is found to exist between resonance frequency estimates and existing lithology. This is well supplemented by the relevant borehole information at some of the selected sites. Additionally, the results, that we obtain from numerical analysis carried out through non-linear site response analysis with the input of acceleration histories from a nearby earthquake from the study area, bear a substantial resemblance with H/V ratio results so far the resonant frequency estimates are concerned.

Within the limitations of H/V ratio as referred in Zare et al. 1999, Zare et al. 2002; Theodulidis et al. 1996; Ghasemi et al. 2009, one can reliably infer the fundamental frequency of a site in a rapid manner provided there exists an impedance contrast between the surface layer and the layer beneath. Besides, as mentioned in the reference (Zare et al., 2002), the H/V ratio technique is able to furnish specially the superficial site conditions; hence, in our case too, we believe that the results yielded by the ambient noise measurements and the following H/V assessments might incorporate the site conditions superficially; not the conditions in depth.

On the basis of these observations delimited by certain restraints, we believe that the present work will help as an initial building block towards detailed microzonation study in Shillong City aided by more analysis such as MASW, GPR etc.

## Data and Resources

The ambient noise data were recorded by Geoscience Division, CSIR-North East Institute of Science & Technology, Jorhat, Assam and the acceleration data was accrued from Civil Engineering Department, Indian Institute of Technology, Guwahati, Assam, India.

## Acknowledgements

We are grateful to Mo.E.S Project , Delhi for their financial support. The Civil Engineering Department of IIT, Guwahati is also acknowledged for providing necessary data.

## References

Badrane S, Bahi L, Jabour N, Brahim AI (2006). Site effects in the city of Rabat (Morocco). J Geophys Eng 3 (3): 207-211

Bard PY (1999) Microtremor measurements, a tool for site affects estimation? Proc 2nd International Symposium Effect of Surface Geology on Seismic Motion. Yokohama, Japan: 1251-1279.

Bardet JP, and Bobita T (2001) Nonlinear Earthquake Site Response Analysis University of Southern California, Department of Civil Engineering.

Baruah S, and Hazarika D (2008) A GIS based tectonic map of Northeastern India, Curr Sci. 95:176-177.

BMTPC, 2003 Vunerability atlas-2nd edn peer group MoH & UPA; seismic zones of India IS, 1983-2002.

Bidyananda, M, and Deomurari MP (2007) Geochronological constraints on the evolution of Meghalaya massif, northeastern India, An ion microprobe study, Curr Sci. 93: 1620-1623.

Bilham, R, and England P (2001) Plateau pop-up in the Great 1897 Assam earthquake. Nature 410: 806-809.

Bonnefoy-Claudet S, Cotton F and Bard PY (2006) The nature of the seismic noise wave field and its implication for site effects studies, a literature review, Earth Sci. Rev. 79: 205–227.

Bonnefoy-Claudet S, Leyton F, Baize S, Berge-Thierry C, Bonilla LF and Campos J (2008) Potentiality of microtremor to evaluate site effects at shallow depths in the deep basin of Santiago de Chile. Proc 14$^{th}$ World Conference on Earthquake Engineering.

Biswas. R., S. Baruah and D. K. Bora, Mapping sediment thickness in Shillong, NER, India, Journal of Earthquakes.

Cara F, Giulio G, and Roveli A (2003) A study on Seismic Noise Variations at Colfiorito, central Italy, Implications for the Use of H/V Spectral Ratios, Geophys. Res. Lett., 30: 1972.

Cara F, Cultera G, Azzara RM, Rubeis V, Giulio G, Giammariano S, Tosi P, and Rovelli A (2008) Microtremor measurements in the city of Palermo, Italy, Analysis of the correlation between local Geology and damage, Bull. Sesimol.. Soc. Am. 98: 1354-1372.

Chattopadhaya N, and Hashimi S (1984) The Sung valley alkaline ultramaffic carbonatite complex, East Khasi Hills district, Meghalaya, Rec Geo Surv India 113: 24-33.

Dubos N, Souriau A, Ponsolles C, and Fels JF (2003) Etude des effets de site dans la ville de Lourdes (Pyrénées, France) par la méthode des rapports spectraux, Bull. Soc. Géol. Fr. 174: 33–44.

Duke CM, Johnsen KE, Larson L, and Engman DC (2007) Effects on site lassification and distance on instrumental indices in the San Fernando earthquake. Rpt. UCLA-ENG-7247. Los Angeles, 50p.

Duval A, Vidal S, Meneroud P, Singer A, Desantis F, Ramos C, G Romero, R Rodriguez, A Pernia, N Reyes, C Griman (2001) Caracas, Venezuela, Site effect determination with micro tremors, Pure Appl. Geophys. 158: 2513-2523

Dunand F, Bard PY, Chatelain JL, Gueguen PH, T Vassail, and MN Farsi (2002) Damping and frequency from Randomdec method applied to the in situ measurements of ambient vibrations, evidence for effective soil structure interaction. Proceedings of 12th European conference on earthquake engineering. London.

Evans, P (1964) The Tectonic Framework of Assam. J Geol Soc India 5: 80-96.

Fernandez LM, and Brandt MBC (2000) The reference spectral noise ratio method to evaluate the seismic response of a site, Soil Dyn Earth. Eng. 20:381-388.

Field, E, and K Jacob (1993) The theoretical response of sedimentary layers to ambient seismic noise, Geophys. Res. Lett. 20(24): 2925-2928.

Garcia-Jerez A, Navarro M, Alcala FJ, Luzon F, Perez Ruiz JA, Enomoto T, Vidal F, and Ocana E (2007) Shallow velocity structure using joint inversion of array and h/v Spectral ratio of ambient noise, The case of Mula Town (SE of Spain), Soil Dyn and Earthquake Eng 27: 907-919.

Ghasemi, H, M Zare, Y Fukushima, F Sinaeian (2009) Applying empirical methods in site classification, using response spectral ratio (H/V), A case study on Iranian Strong motion network (ISMN). Soil Dyn Earth. Eng. 29:121-132.

Gueguen, P, JL Chatelain, B Gullier, and H Yepes (2000) An indication of the soil topmost layer response in Quito (Ecuador) using noise H/V spectral ratio. Soil Dyn Earth. Eng, 19: 127-133.

Guillier, B, JL Chatelain, SB Claudet, and E Haghshenas (2004) Use of ambient noise, From Spectral amplitude variability to H/V Stability. J Earth Eng., 11: 925-942.

Gupta, HK, and VP Singh (1980) Teleseismic P-wave residual Investigations at Shillong, India. Tectonophysics, 66: 19-27.

Haghshenas, E (2005) Condition geotechnique et alea sismique local a Teheran. PhD Thesis, Joseph Fourier University, Grenoble (France), pp. 288.

Hasancebi, N, and R Ulusay (2006) Evaluation of site amplification and site period using different methods for an earthquake-prone settlement in Western Turkey, Eng. Geol. 87: 85-104.

Huang, CS, and CH Yeh (1999) Some properties of the Randomdec signatures. Mechanical Systems and Signal Processing 13:491-507.

Imai, T, and K Tonouchi (1982) Correlation of N-value with S-wave velocity and Shear Modulus. In, Proceedings 2nd European Symposium on Penetration Testing Amsterdam, 57-72.

Kalita, BC (1998) Ground water prospects of Shillong Urban Aglomerate. Unpublished report. CGWB.

Khatri, KN, R Chander, S Mukhopadhyay, V Sriram, and KN Khanal (1992) A model of active tectonics in the Shillong Massif region, Himalayan Orogen and Global Tectonics.

Komatitsch, D, and JP Vilotte (1998) The spectral-element method, an efficient tool to simulate the seismic response of 2D and 3D geological structures, Bull. Sesim. Soc. Am. 88(2): 368–392.

Konno, K, and T Ohmachi (1998) Ground motion characteristics estimated from spectral ratio between horizontal and vertical components of microtremor. Bull. Sesim. Soc. Am. 88:228-241.

Lebrun, B, D Hatzfeld, and PY Bard (2001) A site effect study in urban area, experimental results in Grenoble (France). Pure Appl. Geophys 158: 2543-2557.

Lermo, J, and FJ Chavez-Garcia (1994) Are microtremors useful in site response evaluation? Bull. Seism. Soc. Am. 84: 1350–1364.

Lombardo, G, Rigano, R, 2007 Local seismic response in Catania (Italy), A test area in the Northern part of the Town. Eng Geo 94: 38-49.

Mikhailova, NN, and FF Aptikaev (1996) Some correlation relations between parameters of seismic motions. J. Earth. Prediction Res., 5:257-267.

Mundepi AK, Lindholm C, Kamal C (2009) Soft soil mapping using horizontal to vertical ratio (HVSR) for Seismic Hazard Assessment of Chandigarh City in Himalayan Foothills, North India. Jour. Geol. Soc. India., 74(5): 551-558.

Nandy DR (2001) Geodynamics of Northeastern India and the adjoining region. ACB publications, Calcutta, pp 209.

Nakamura Y (1989) A method for dynamic characteristics estimation of Sub surface using microtremor on the surface. Railway Technical Research Institute Report, 3025-3033.

Oldham RD (1899) Report on the great earthquake of 12[th] June 1897. Mem. Geol. Surv. India 29: 1-379.

Ouchi T, Lin A, and Maruyama T (2001) The 1999 Chi-Chi (Taiwan) earthquake, earthquake fault and strong motions, Bull Seism. Soc. Am. 91: 966–976.

Panou A, Theodulidis N, Hatzidimitriou P, Savvaidis A, Papazachos C (2005) Reliability of ambient noise horizontal to vertical spectral ratio in urban environments, the case of Thessaloniki City (Northern Greece) Pure Appl. Geophys 162:891-912.

Poddar MC (1950) Preliminary report of the Assam earthquake of 15[th] August 1950. Geol. Surv. India Bull. Ser B (2): 1-40.

Rao JM, and Rao GVSP (2008) Geology, Geochemistry and Palaeomagnetic study of Cretaceous Mafic Dykes of Shillong Plateau and Their Evolutionary History Indian Dykes, Geochemistry, Geophysics and Geomorphology 589-607.

Rao JM and Rao GVSP (2009) Precambrian mafic magmatism of Shillong plateau, Meghalaya and their evolutionary history, J. Geol. Soc. India 73:143-152.

Richter CF (1958) Elementary Seismology. W.H. Freeman & Co San Francisco; pp768.

Sar, SN (1973) An interim report on ground water exploration in the Greater Shillong area, Khasi Hills District, Meghalaya, Memo report, Central Ground Water Board

SESAME (2004) Guidelines for the implementation of the H/V spectral ratio technique on ambient vibrations measurements, processing and interpretation. SESAME European research project WP12-Deliverable D23.12, December.

Souriau, A, A Roulle, and C Ponsolles (2007) Site Effects in the City of Lourdes, France, from H/V Measurements: Implications for Seismic-Risk Evaluation. Bull. Seism. Soc. Am., 97(6): 2118–2136.

Srinivassan P, Sen S, and Bandopadhaya PC (1996) Study of variation of Paleocene-Eocene sediments in the shield areas of Shillong Plateau. Rec Geol Surv India 129: 77-78.

Theodulidis N, Bard PY, Archuleta R, and Bouchon M (1996) Horizontal to Vertical Spectral ratio and geological Conditions, The case of Garner valley downhole array in southern California. Bull. Seism. Soc. Am. 86(2): 306-319.

Tillotson E (1953) The Great Assam earthquake of 1950, the completion of papers on the Assam Earthquake of August 15, 1950, compiled by MB Ramachandra Rao, 94-96

Tokeshi, JC and Sugimura Y (1998) On the estimation of the natural period of the round using simulated microtremors. In, 2nd International Symposium on the Effects of Surface Geology on Seismic Motion, Yokohama, Japan 2: 651–664.

Wakamatsu K, and Yasui Y (1996) Possibility of estimation for amplification characteristics of soil deposits based on ratio of horizontal to vertical spectra of microtremors. In, 11th World Conference on Earthquake Engineering, Acapulco, Mexico.

Zare M, Bard PY, and Ghafory-Ashtiany M (1999) Site characterizations for the Iranian strong motion network. Soil. Dyn. Earth. Eng. 18(2): 101–23.

Zare M, and Bard PY (2002) Strong motion dataset of Turkey: data processing and site classification. Soil Dyn. Earth. Eng. 22: 708-719.

**Table 1** Epicentral parameters of the events

| Date dd-mm-yy | Origin Time hh:mm:ss | Latitude $^0$E | Longitude $^0$N | Depth (m) | Mag |
|---|---|---|---|---|---|

| 19-08-09 | 10:45:13.87 | 26.56 | 92.47 | 10 | 5.1 |

**Table 2** Ground motion parameters estimated through NERA at each borehole site.

| Borehole ID | Spectral Acceleration g (cm/s$^2$) | Maximum Spectral Velocity(cm/s) | Fundamental Frequency (Hz) | Av. Shear Wave velocity (m/s) |
|---|---|---|---|---|
| Pynthorumkhrah | 0.6 | 7.1 | 5.1 | 393 |
| Umpling | 0.4 | 8.6 | 2.6 | 362 |
| Arbuthnot Road | 0.5 | 15.9 | 4.6 | 332 |
| East of Pinewood Hotel | 0.2 | 6.3 | 5.8 | 356 |
| Sawlad | 0.5 | 16.6 | 5.5 | 346 |
| Neigrims | 0.4 | 11.0 | 5.0 | 384 |
| Near 13 A Tank | 0.2 | 6.3 | 8.7 | 323 |
| B.S.N.L | 0.4 | 11.8 | 5.1 | 402 |
| Near Pay Ward | 0.3 | 6.4 | 12.5 | 380 |
| Maw pat | 0.4 | 10.6 | 4.5 | 425 |

**Table 3** Comparison of fundamental frequencies estimated from NERA and H/V ratios

| Borehole IDS. | 4.Fundamental* Frequency (Hz) | 3Fundamental# Frequency (Hz) |
|---|---|---|
| Neigrims | 5.08 | 5.40 |
| Near 13 A Tank | 8.79 | 1.20 |
| Near Pay Ward | 12.50 | 7.10 |
| Umpling | 2.59 | 2.70 |
| Saw lad | 5.55 | 4.47 |
| Pynthoumkhrah | 5.08 | 6.00 |
| Arbhutnot Road | 4.59 | 6.05 |
| East of Pine wood Hotel | 5.08 | 3.50 |
| Mawpat | 4.45 | 4.40 |
| B.S.N.L | 5.08 | 3.43 |

\* Average Fundamental frequency computed from NERA through simulation of input acceleration histories.
\# Fundamental frequency computed from H/V ratio estimate of the ambient noise records.

**Figure Captions:**

**Figure 1** Geological map of Shillong City

**Figure 2** Map showing major tectonic features of Shillong plateau after Baruah and Hazarika (2008). The inset map shows the study region. BS-Barapani Shear Zone; SF- Samin Fault; DT- Dapsi Thrust; Du F- Dudhnoi Fault; OF- Oldham Fault; CF-Chedrang Fault; BL- Bomdila Lineament.

**Figure 3** Ambient noise location points along with borehole locations. Filled circles represent the location of the sites where ambient noise recordings, represented by no's as well are made. Filled triangles refer to the location of the boreholes. AB and CD are the observed profiles.

**Figure 4** Damping results from Randomdec Method to check the origin of the H/V estimate of peak frequency for a particular site.

**Figure 5** Higher fundamental frequencies from H/V ratio. The grey vertical line represents the peak of the H/V ratio corresponding to the resonant frequency. The dashed line indicates the standard deviation whereas the average H/V ratio is indicated by the solid line.

**Figure 6** H/V ratio results encompassed by low frequency peaks.

**Figure 7** Contour plots showing the distribution of fundamental frequencies computed from H/V ratio for the entire Shillong Area.

**Figure 8** (a) Variation of fundamental frequencies along AB profile. Along the vertical axis, fundamental frequency is plotted and distance is provided along horizontal axis. The color scale appearing at the right indicates the variation of amplitudes corresponding to the computed fundamental frequencies.

(b) Plot of ambient noise sites falling in the profile AB. Along X and Y axis; distances in meters among the sites are shown. The color scale implies the corresponding site –specific H/V peak frequencies.

**Figure 9** Variation of fundamental frequencies along CD profile. Along the vertical axis, fundamental frequency is plotted and distance is provided along horizontal axis. The colour scale appearing at the right indicates the variation of amplitudes corresponding to the computed fundamental frequencies. (b) Plot of ambient noise sites falling in the profile CD. Along X and Y axis; distances in meters among the sites are shown. The colour scale implies the corresponding site –specific H/V peak frequencies.

**Figure 10** Variation of fundamental frequencies along AB profile. The figures in the boxes indicate the ambient noise sites alongside of AB profile. The bold vertical lines indicate the location of the boreholes (b) Depiction of litho-section along AB profile with corresponding

topographical variation. Here, depth is projected vertically downward with an equal interval of 50m

**Figure 11** (a) Variation of fundamental frequencies along CD profile. The figures in the boxes indicate the ambient noise sites alongside of CD profile. The bold vertical lines indicate the location of the boreholes (b) Depiction of litho-section along CD profile with corresponding topographical variation. Here, depth is projected vertically downward with an equal interval of 50m

**Figure 12.** Estimated parameters from non-linear earthquake site response analysis. (a) Input acceleration time history of 19-08-09 event. (b) Simulated acceleration history. (c) Estimated fundamental frequency. (d) Borehole log data at one of the borehole site. (e) Shear wave-velocity profile of three borehole sites viz. Arbhutnot Road. East of Pinewood Hotel (abbreviated as East Hotel) & Pynthorumkhrah. (f) Spectral acceleration. (g) Spectral Relative Velocity.

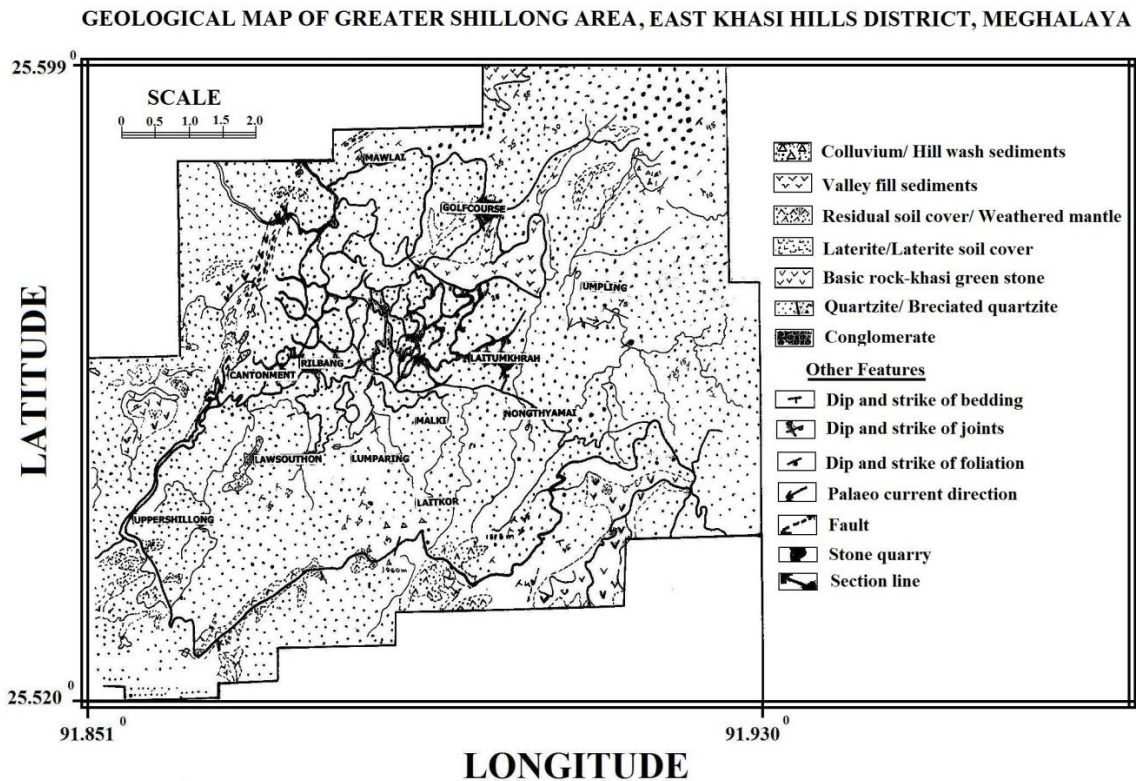

**Figure 1** Geological map of Shillong City

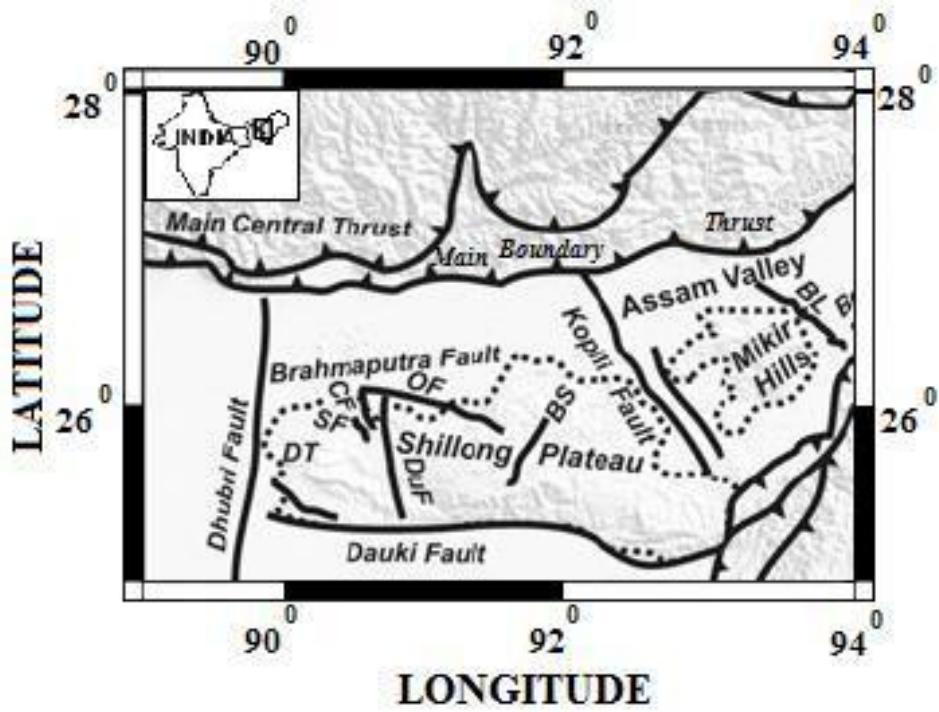

**Figure 2** Map showing major tectonic features of Shillong plateau after Baruah and Hazarika (2008). The inset map shows the study region. BS-Barapani Shear Zone; SF- Samin Fault; DT- Dapsi Thrust; Du F- Dudhnoi Fault; OF- Oldham Fault; CF-Chedrang Fault; BL- Bomdila Lineament.

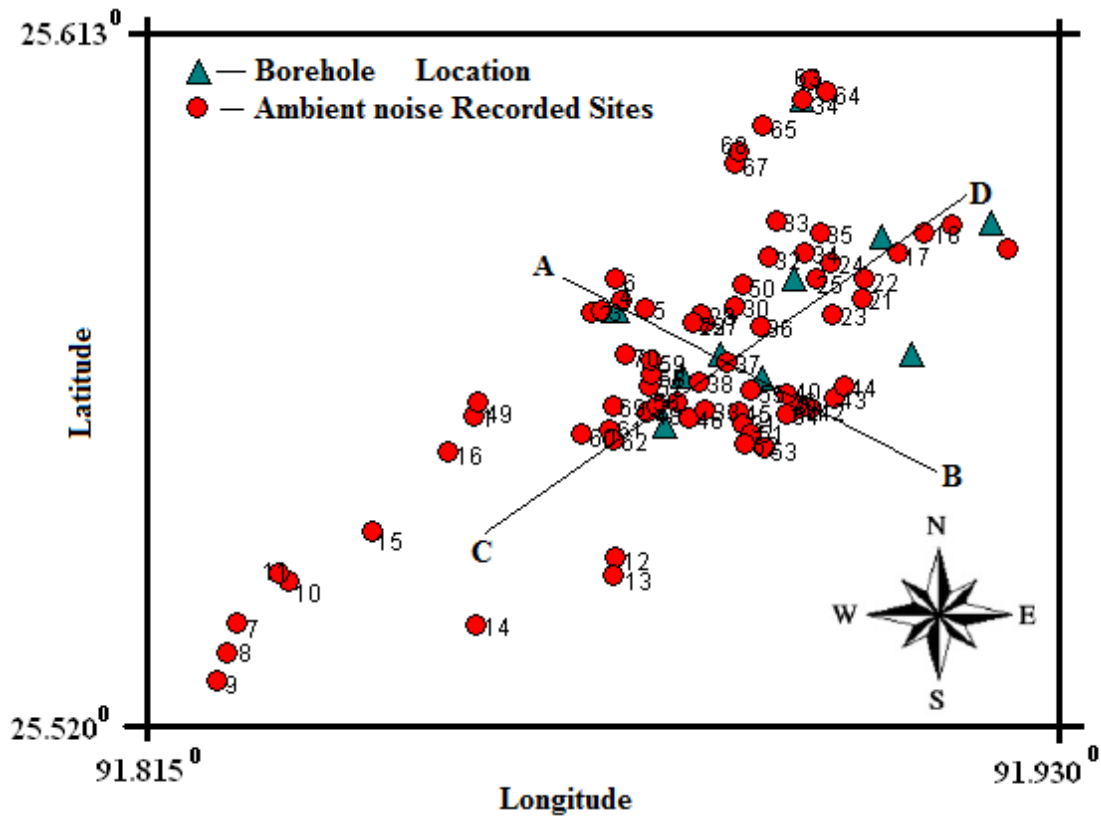

**Figure 3** Ambient noise location points along with borehole locations. Filled circles represent the location of the sites where ambient noise recordings, represented by no's as well are made. Filled triangles refer to the location of the boreholes. AB and CD are the observed profiles.

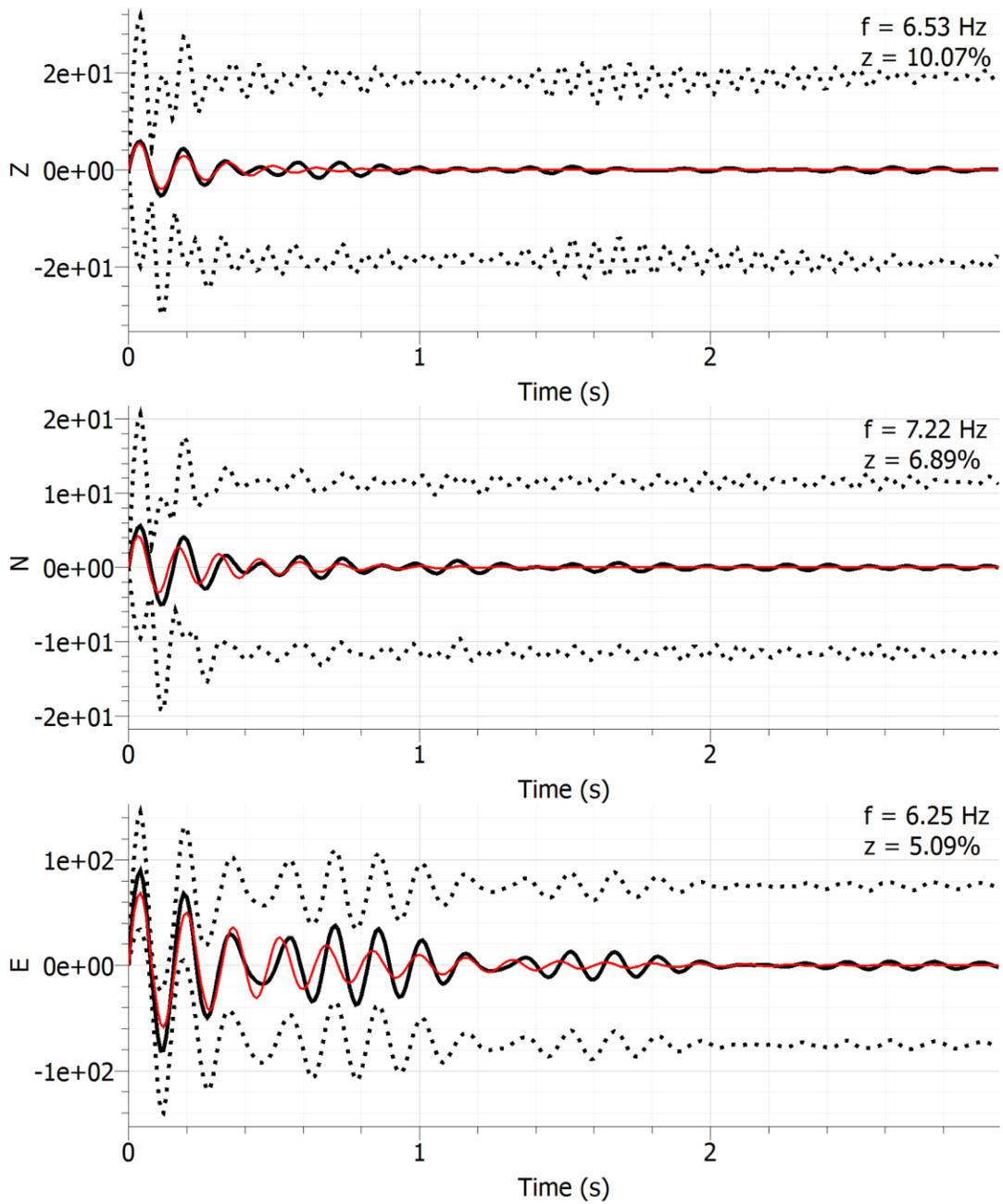

**Figure 4** Damping results from Randomdec Method to check the origin of the H/V estimate of peak frequency for a particular site.

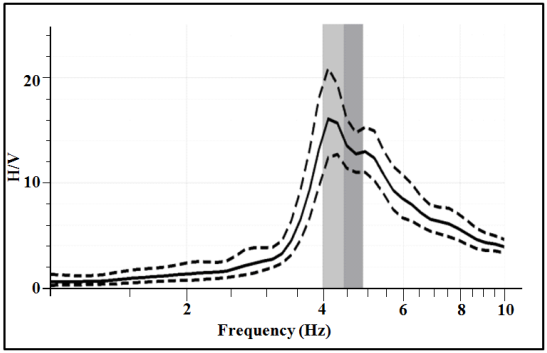 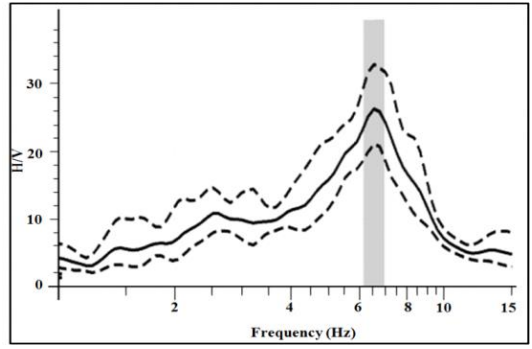

(a)  (b)

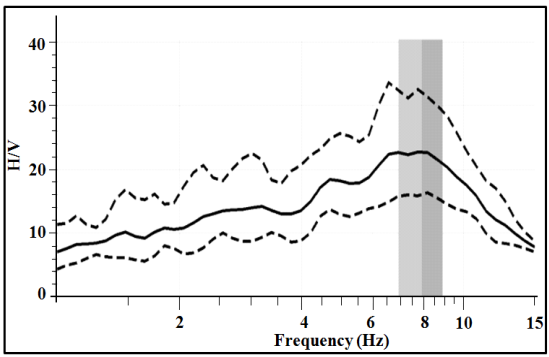 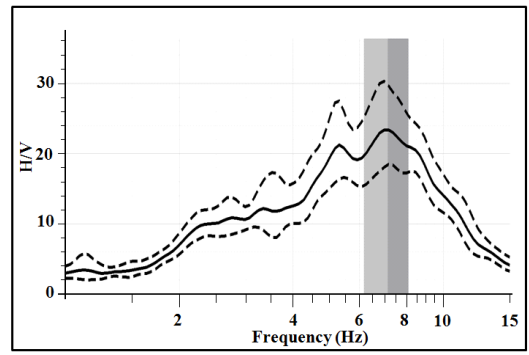

(c)  (d)

**Figure 5** Higher fundamental frequencies from H/V ratio. The grey vertical line represents the peak of the H/V ratio corresponding to the resonant frequency. The dashed line indicates the standard deviation whereas the average H/V ratio is indicated by the solid line. (a) Site No. 30 (b) Site No. 42     (c) Site No. 13     (d) Site No. 64

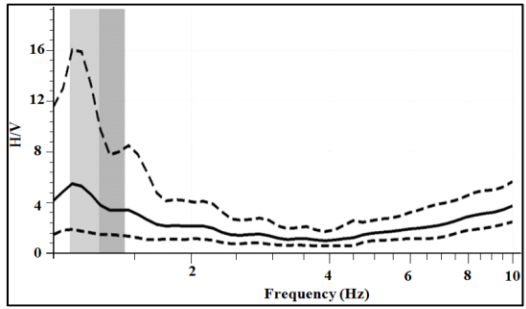
(a)

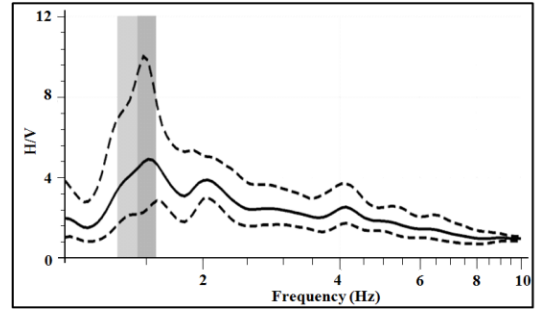
(b)

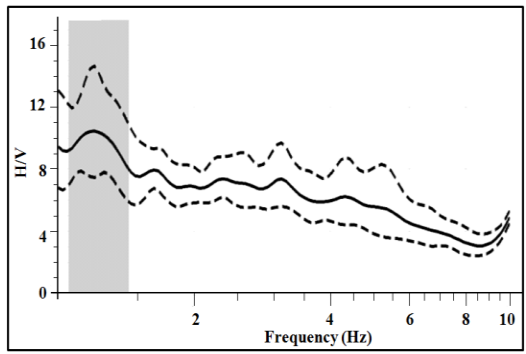
(c)

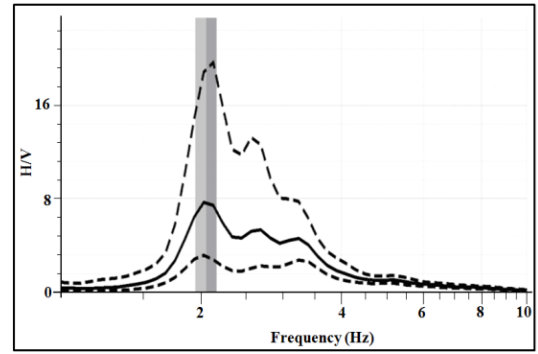
(d)

**Figure 6** H/V ratio results encompassed by low frequency peaks. The grey vertical line represents the peak of the H/V ratio corresponding to the resonant frequency. The dashed line indicates the standard deviation whereas the average H/V ratio is indicated by the solid line. (a) Site No. 27 (b) Site No. 54 (c) Site No. 50 (d) Site No. 21

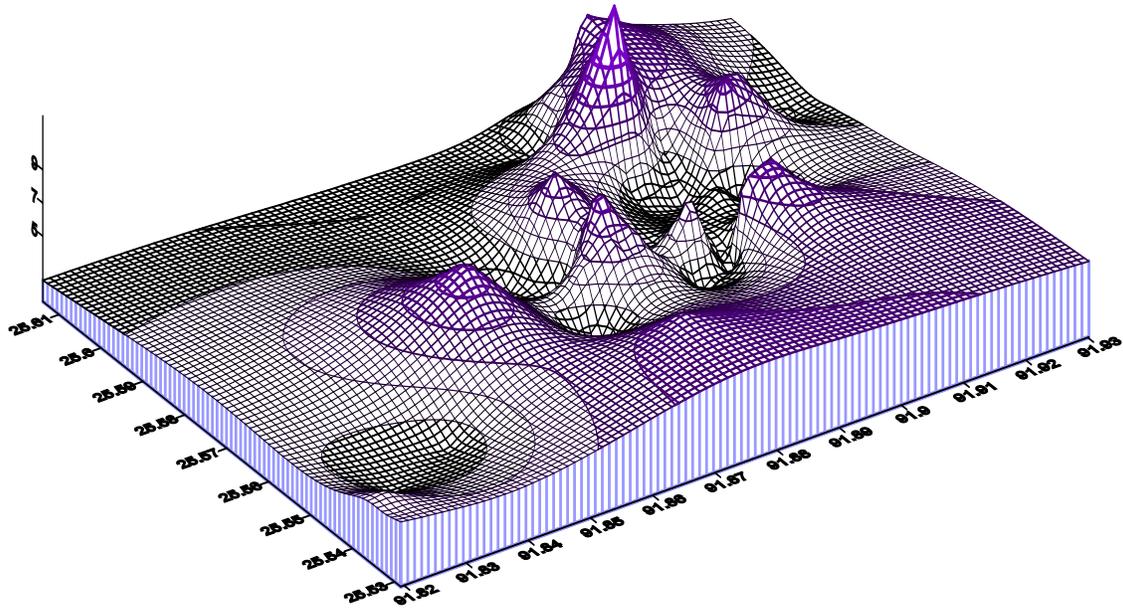

**Figure 7** Contour plots showing the distribution of fundamental frequencies computed from H/V ratio for the entire Shillong Area.

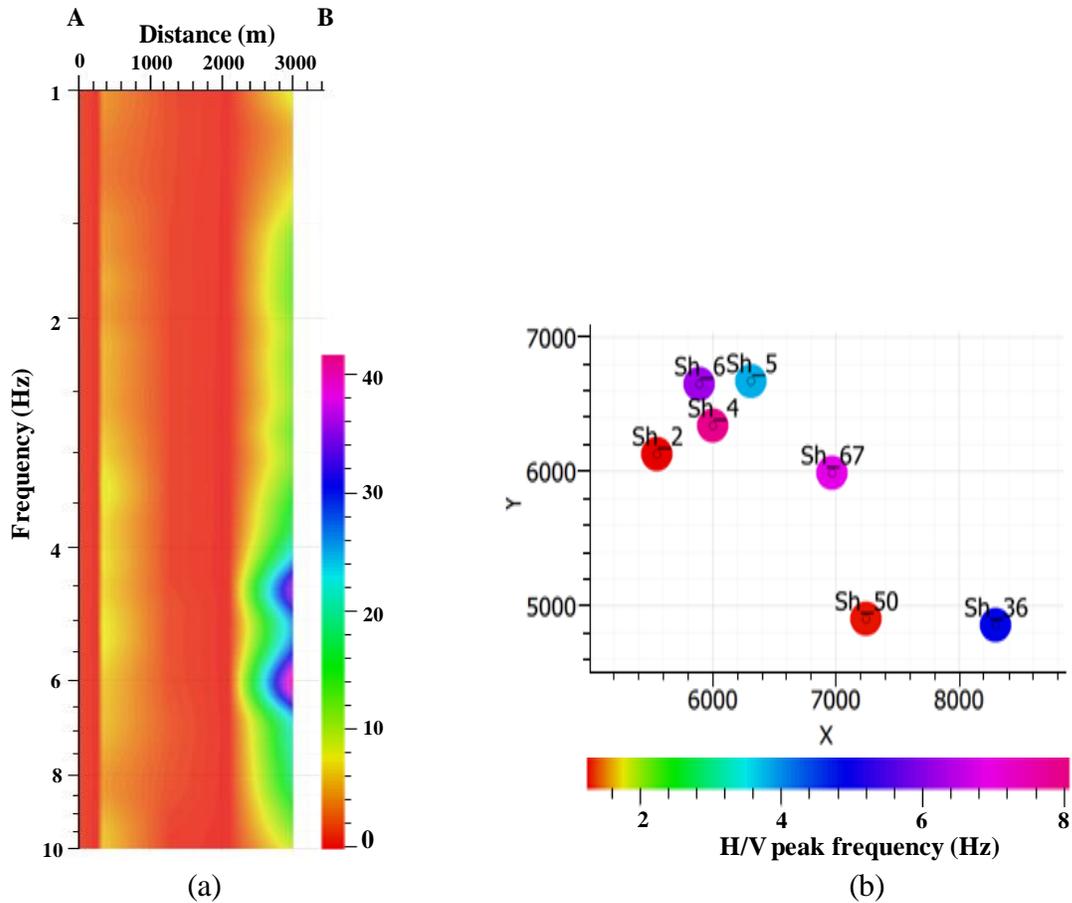

**Figure 8** (a) Variation of fundamental frequencies along AB profile. Along the vertical axis, fundamental frequency is plotted and distance is provided along horizontal axis. The color scale appearing at the right indicates the variation of amplitudes corresponding to the computed fundamental frequencies. (b) Plot of ambient noise sites falling in the profile AB. Along X and Y axis; distances in meters among the sites are shown. The color scale implies the corresponding site –specific H/V peak frequencies.

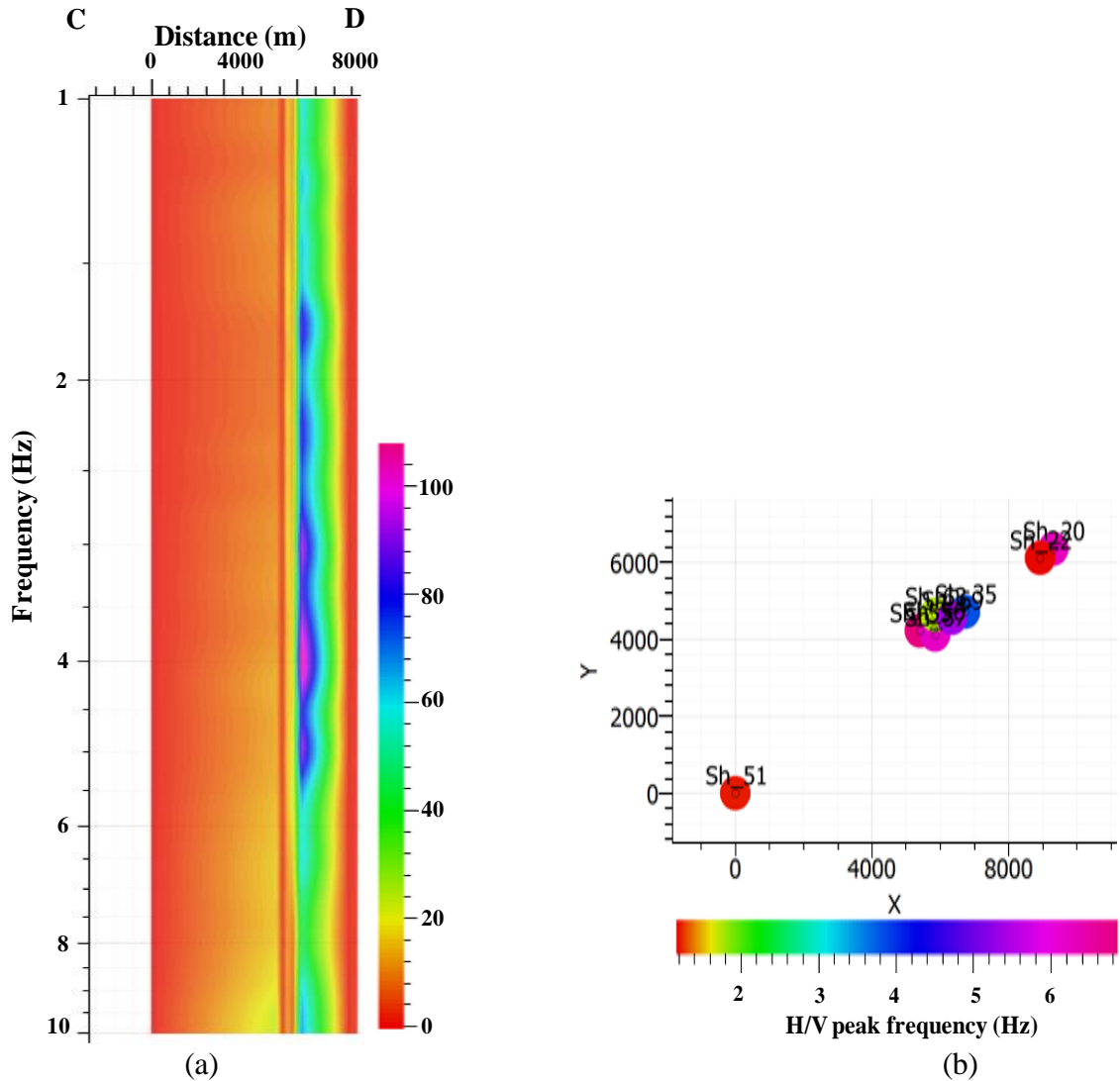

**Figure 9** Variation of fundamental frequencies along CD profile. Along the vertical axis, fundamental frequency is plotted and distance is provided along horizontal axis. The colour scale appearing at the right indicates the variation of amplitudes corresponding to the computed fundamental frequencies. (b) Plot of ambient noise sites falling in the profile CD. Along X and Y axis; distances in meters among the sites are shown. The colour scale implies the corresponding site –specific H/V peak frequencies.

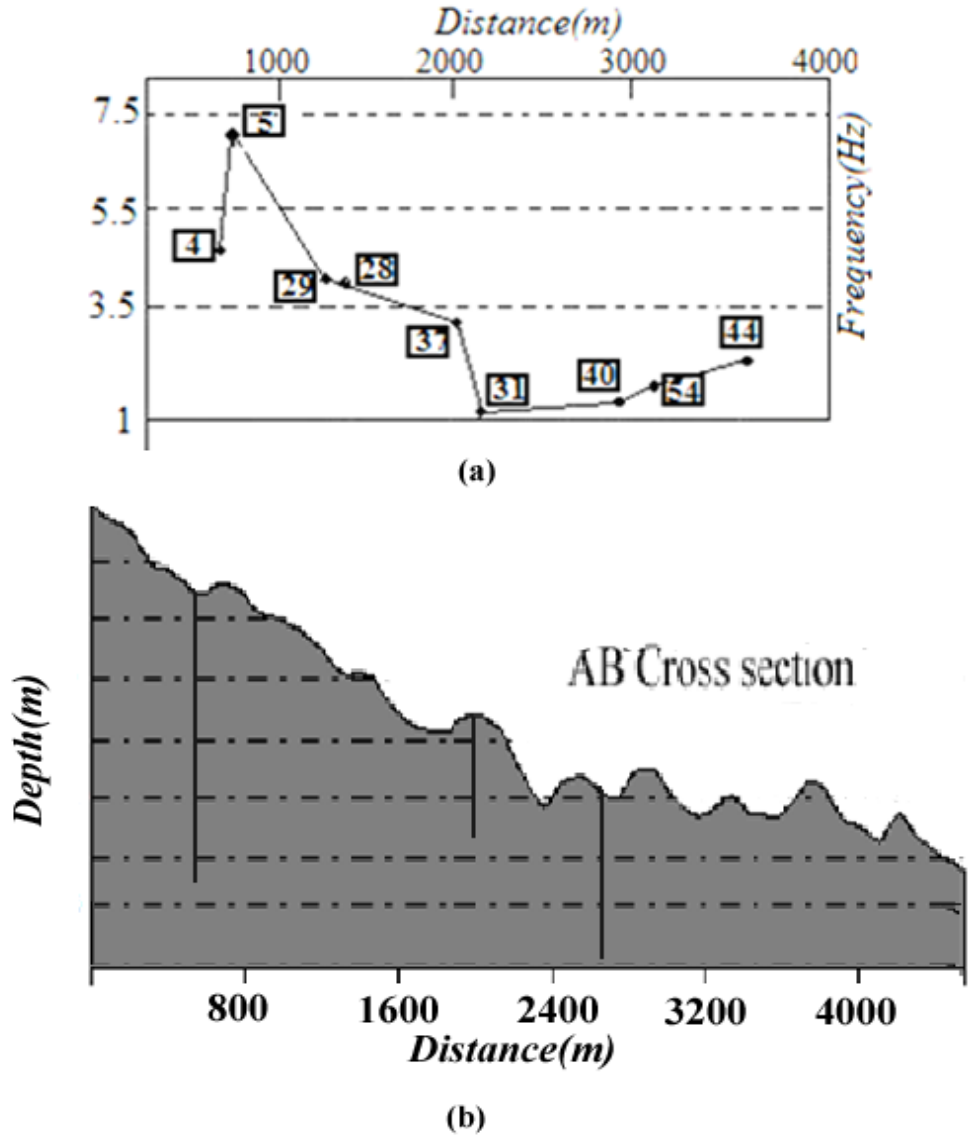

**Figure 10** Variation of fundamental frequencies along AB profile. The figures in the boxes indicate the ambient noise sites alongside of AB profile. The bold vertical lines indicate the location of the boreholes (b) Depiction of litho-section along AB profile with corresponding topographical variation. Here, depth is projected vertically downward with an equal interval of 50m

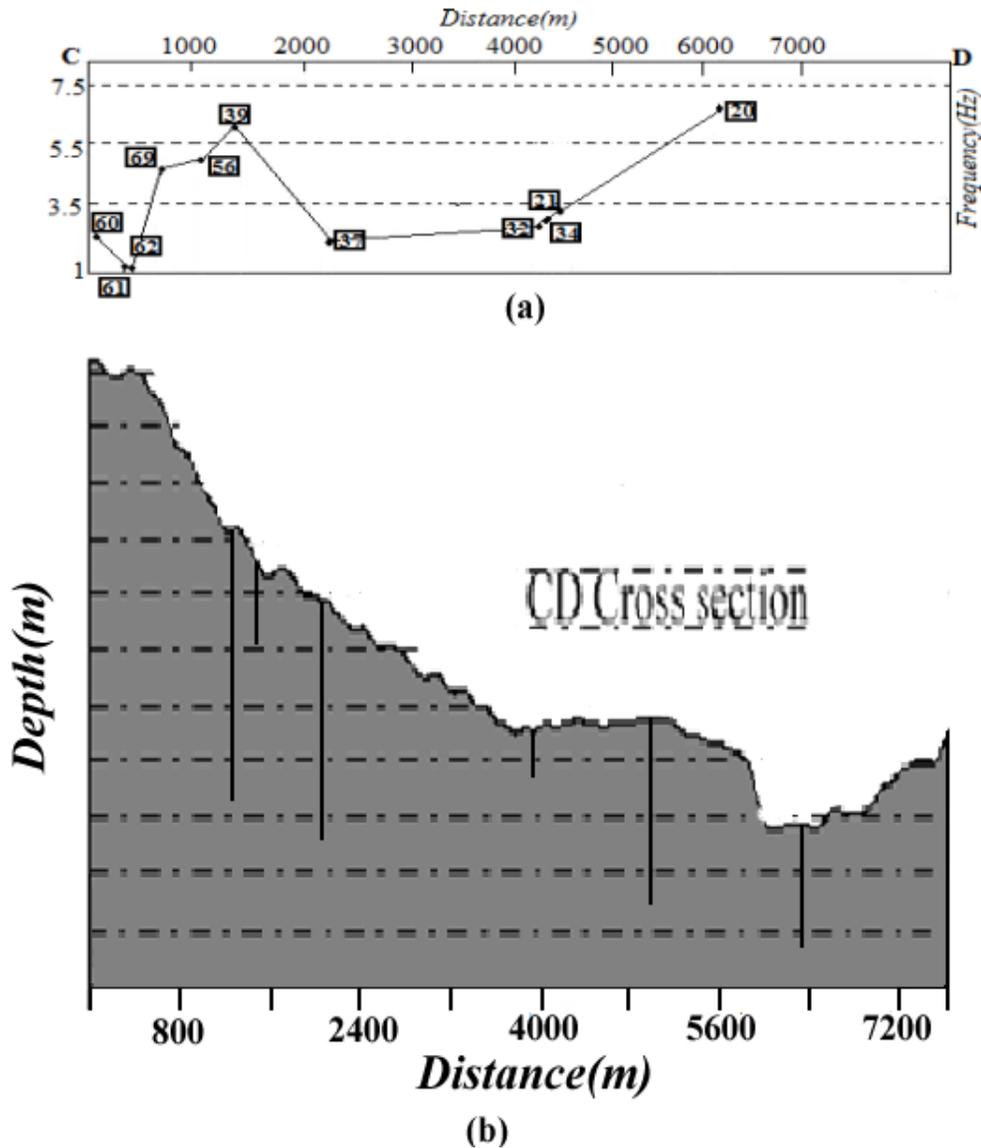

**Figure 11** (a) Variation of fundamental frequencies along CD profile. The figures in the boxes indicate the ambient noise sites alongside of CD profile. The bold vertical lines indicate the location of the boreholes (b) Depiction of litho-section along CD profile with corresponding topographical variation. Here, depth is projected vertically downward with an equal interval of 50m

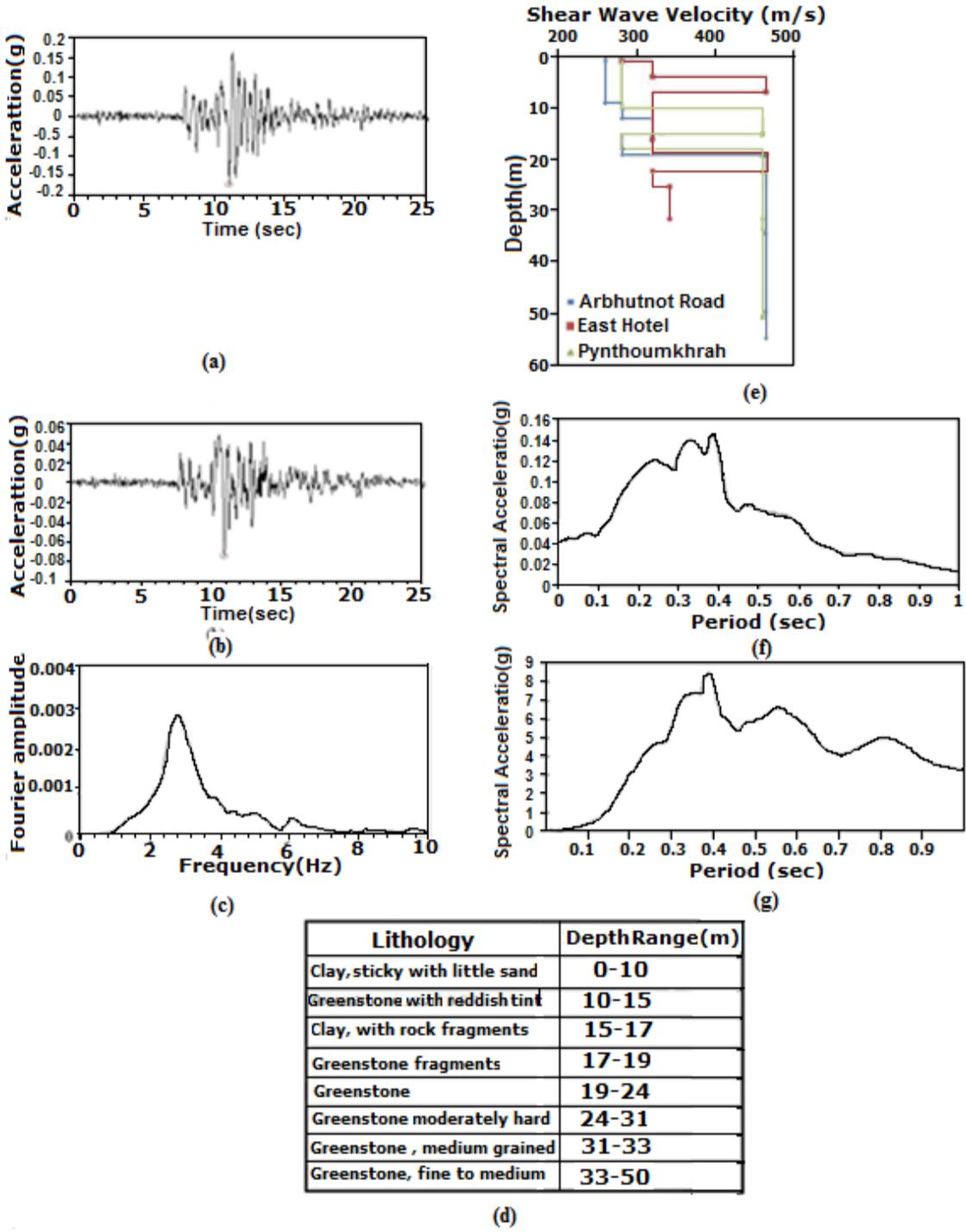

**Figure 12** Estimated parameters from non-linear earthquake site response analysis. (a) Input acceleration time history of 19-08-09 event. (b) Simulated acceleration history. (c) Estimated fundamental frequency. (d) Borehole log data at one of the borehole site. (e) Shear wave-velocity profile of three borehole sites viz. Arbhutnot Road. East of Pinewood Hotel (abbreviated as East Hotel) & Pynthorumkhrah. (f) Spectral acceleration. (g) Spectral Relative Velocity.